 \documentclass[twocolumn,pre,aps,showpacs,preprintnumbers,amsmath,amssymb,superscriptaddress]{revtex4}
\usepackage{epstopdf}
\usepackage{graphicx}
\usepackage{graphicx,epsf} 
\usepackage{dcolumn}
\usepackage{bm}
\usepackage{latexsym}
\newcommand{\kav}{\langle k \rangle}
\newcommand{\xsize}{\epsfxsize=9 cm}

\begin{document}

\title{Social consensus through the influence of committed minorities}

\author{J. Xie}
\affiliation{Dept. of Computer Science, Rensselaer Polytechnic Institute, 110 8th Street, Troy, NY 12180} 

\author{S. Sreenivasan \footnote{Corresponding author: sreens@rpi.edu} } 
\affiliation{Dept. of Computer Science, Rensselaer Polytechnic Institute, 110 8th Street, Troy, NY 12180} 
\affiliation{Dept. of Physics, Rensselaer Polytechnic Institute, 110 8th Street, Troy, NY 12180} 

\author{G. Korniss}
\affiliation{Dept. of Physics, Rensselaer Polytechnic Institute, 110 8th Street, Troy, NY 12180}

\author{W. Zhang}
\affiliation{Dept. of Mathematics, Rensselaer Polytechnic Institute, 110 8th Street, Troy, NY 12180} 

\author{C. Lim}
\affiliation{Dept. of Mathematics, Rensselaer Polytechnic Institute, 110 8th Street, Troy, NY 12180} 
 
\author{B. K. Szymanski}
\affiliation{Dept. of Computer Science, Rensselaer Polytechnic Institute, 110 8th Street, Troy, NY 12180}

\begin{abstract}
We show how the prevailing majority opinion in a population can be rapidly reversed by a small fraction $p$ of randomly distributed {\it committed} agents who consistently proselytize the opposing opinion and are immune to influence. Specifically, we show that when the committed fraction grows beyond a critical value  $p_c \approx 10 \%$, there is a dramatic decrease in the time, $T_c$, taken for the entire population to adopt the committed opinion. In particular, for complete graphs we show that  when $p < p_c$, $T_c \sim \exp(\alpha(p)N)$, while for $p > p_c$, $T_c \sim \ln N$. We conclude with simulation results for Erd\H{o}s-R\'enyi random graphs and scale-free networks which show qualitatively similar behavior. 
\end{abstract}

\maketitle

\section{Introduction}
Human behavior is profoundly affected by the influenceability of individuals and the social networks that link them together. Well before the proliferation of online social networking, offline or interpersonal social networks have been acknowledged as a major factor in determining how societies move towards consensus in the adoption of ideologies, traditions and attitudes \cite{Harary1959,Friedkin1990}. As a result, the dynamics of social influence has been heavily studied in sociological, physics and computer science literature \cite{Schelling1978,Castellano_review2009, Kempe2003, Galam1999, Galam2008}. In the sociological context, work on {\it diffusion of innovations} has emphasized how individuals adopt new states in behavior, opinion or consumption through the influence of their neighbors. Commonly used models for this process include the threshold model \cite{Granovetter1978} and the Bass model \cite{Bass1969}. A key feature in both these models is that once an individual adopts the new state, his state remains unchanged at all subsequent times.  While appropriate for modeling the diffusion of innovation where investment in a new idea comes at a cost, these models are less suited to studying the dynamics of competing opinions where switching one's state has little overhead. 

Here we address the latter case. From among the vast repertoire of models in statistical physics and mathematical sociology, we focus on one which is a 2-opinion variant \cite{Castello2009} of the Naming Game (NG) \cite{Steels1995,Baronchelli2006,DallAsta2006,Lu2008,Baronchelli2011} and that we refer to as the {\it binary agreement model}. The evolution of the system in this model takes place through the usual NG dynamics, wherein at each simulation time step, a randomly chosen speaker voices a random opinion from his list to a randomly chosen neighbor, designated the listener. If the listener has the spoken opinion in his list, both speaker and listener retain only that opinion, else the listener adds the spoken opinion to his list (see Table~1). The order of selecting speakers and listeners is known to influence the dynamics, and we stick to choosing the speaker first, followed by the listener. 

It serves to point out that an important difference between the binary agreement model and the predominantly used opinion dynamics models \cite{Castellano_review2009, Galam1999, Stauffer2002, Krapivsky2003,Sood2005} is that an agent is allowed to possess both opinions simultaneously in the former, and this significantly alters the time required to attain consensus starting from uniform initial conditions. Numerical studies in \cite{Castello2009} have shown that for the binary agreement model on a complete graph, starting from an initial condition where each agent randomly adopts one of the two opinions with equal probability, the system reaches consensus in time $T_c \sim \ln N$ (in contrast, for example, to $T_c \sim N$ for the voter model). Here $N$ is the number of nodes in the network, and unit time consists of $N$ speaker-listener interactions. The binary agreement model is well suited to understanding how opinions, perceptions or behaviors of individuals are altered through social interactions specifically in situations where the cost associated with changing one's opinion is low \cite{Uzzi2011}, or where changes in state are not deliberate or calculated, but unconscious \cite{Christakis2007}. Furthermore, by its very definition, the binary agreement model may be applicable to situations where agents while trying to influence others, simultaneously also have a desire to reach global consensus \cite{Kearns2009}. 

Another merit of the binary agreement dynamics in modeling social opinion change seems worth mentioning. Two state epidemic-like models of social ``contagion" (examples in \cite{Watts2007}) suffer from the drawback that the rules governing the conversion of a node from a given state to the other are not symmetric for the two states. In contrast, in the binary agreement model, both singular opinion states are treated symmetrically in their susceptibility to change. 

Here we study the evolution of opinions in the binary agreement model starting from an initial state where all agents adopt a given opinion $B$, except for a finite fraction $p$ of the total number of agents who are {\it committed agents} and have state $A$.  Committed agents, introduced previously in \cite{Lu2009}, are defined as nodes that can influence other nodes to alter their state through the usual prescribed rules, but which themselves are immune to influence. In the presence of committed agents adopting state $A$, the only absorbing fixed point of the system is the consensus state where all influenceable nodes adopt opinion $A$ - the opinion of the committed agents. The question that we specifically ask is:  how does the consensus time vary with the size of the committed fraction? More generally, our work addresses the conditions under which an inflexible set of minority opinion holders can win over the rest of the population. 

The effect of having un-influencable agents has been considered to some extent in prior studies.  Biswas et al. \cite{Biswas2009} considered for two state opinion dynamics models in one dimension, the case where some individuals are ``rigid" in both segments of the population, and studied the time evolution of the magnetization and the fraction of domain walls in the system. Mobilia et. al. \cite{Mobilia2007} considered the case of the voter model with some fraction of spins representing ``zealots" who never change their state, and studied the magnetization distribution of the system on the complete graph, and in one and two dimensions. Our study differs from these not only due to the particular model of opinion dynamics, but also in its explicit consideration of different network topologies and of finite size networks, specifically deriving how consensus times scale with network size for the particular case of the complete graph. Furthermore, the above mentioned studies do not explicitly consider the initial state that we care about - one where the entire minority set is un-influencable. A notable exception to the latter is the study by Galam and Jacobs \cite{Galam2007} in which the authors considered the case of ``inflexibles" in a two state model of opinion dynamics with opinion updates obeying a majority rule. While this study provides several useful insights and is certainly the seminal quantitive attempt at understanding the effect of committed minorities, its analysis is restricted to the mean-field case, and has no explicit consideration of consensus times for finite systems.

\section{Complete graphs}\label{sec:CG}
\subsection{Infinite network size limit}
We start along similar lines as \cite{Galam2007} by considering the case where the social network connecting agents is a complete graph with the size of the network $N \to \infty$. We designate the densities of uncommitted nodes in states $A,B$ as $n_A, n_B$ respectively. Consequently, the density of nodes in the mixed state $AB$ is $n_{AB}= 1 - p - n_A - n_B$, where $p$ is the fraction of the total number of nodes, that are committed. Neglecting correlations between nodes, and fluctuations, one can write the following rate equations for the evolution of densities:

\begin{eqnarray}
\frac{dn_A}{dt} &=& -n_A n_B  +  n_{AB}^2  + n_{AB} n_A + \frac{3}{2} p n_{AB}  \nonumber  \\
\frac{dn_B}{dt} &=& -n_A n_B  +  n_{AB}^2  + n_{AB} n_B - p n_B 
\label{MF}
\end{eqnarray}

The terms in these equations are obtained by considering all interactions which increase (decrease) the density of agents in a particular state and computing the probability of that interaction occurring. Table~1 lists all possible interactions. As an example, the probability of the interaction listed in row eight is equal to the probability that a node in state AB is chosen as speaker and a node in state B is chosen as listener ($n_{AB}n_{A}$) times the probability that the speaker voices opinion $A$ ($\frac{1}{2}$).

\begin{table}
\centering
\begin{tabular}{|c|c|} \hline
Before interaction & After interaction \\ \hline
A $\overset{A}{\rightarrow}$   A & A - A \\ \hline
A $\overset{A}{\rightarrow}$   B & A - AB\\ \hline
A  $\overset{A}{\rightarrow}$   AB & A - A\\ \hline
B  $\overset{B}{\rightarrow}$   A & B - AB \\ \hline
B  $\overset{B}{\rightarrow}$   B & B - B\\ \hline
B  $\overset{B}{\rightarrow}$   AB & B - B\\ \hline
AB  $\overset{A}{\rightarrow}$   A & A - A \\ \hline
AB  $\overset{A}{\rightarrow}$   B & AB - AB\\ \hline
AB  $\overset{A}{\rightarrow}$   AB & A - A\\ \hline
AB  $\overset{B}{\rightarrow}$   A & AB - AB \\ \hline
AB  $\overset{B}{\rightarrow}$   B & B - B\\ \hline
AB  $\overset{B}{\rightarrow}$   AB & B - B\\ \hline
\end{tabular}
\label{interactions}
\caption{Shown here are the possible interactions in the binary agreement model. Nodes can possess opinion $A$, $B$ or $AB$, and opinion updates occur through repeated selection of speaker-listener pairs. Shown in the left column are the opinions of the speaker (first)and listener (second) before the interaction, and the opinion voiced by the speaker during the interaction is shown above the arrow. The column on right shows the states of the speaker-listener pair after the interaction.}
\end{table}

\begin{figure*}[!htbp]
\centerline{
\epsfxsize=16 cm
\epsfclipon
\epsfbox{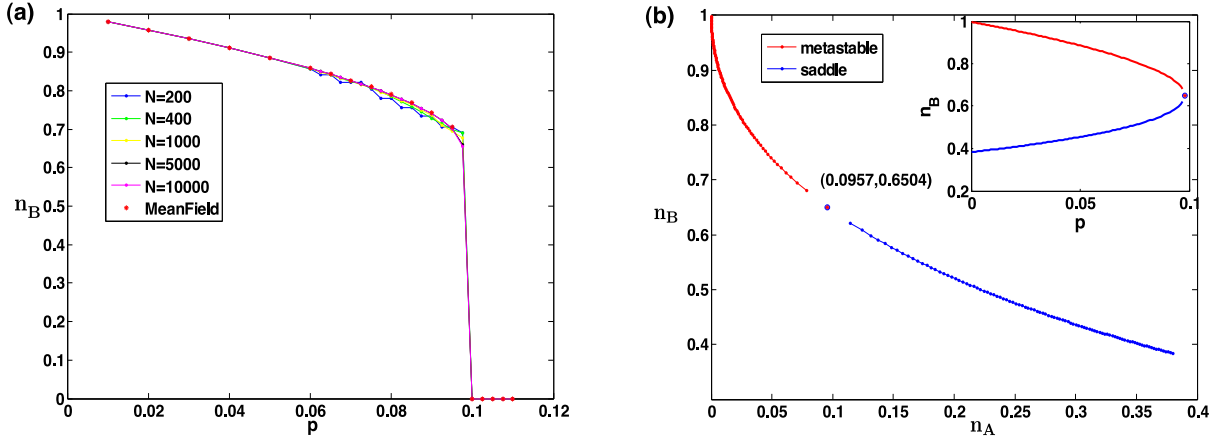}
}
\caption{(a) The steady state density $n_B$ of nodes in state $B$ as a function of committed fraction $p$ for complete graphs of different sizes, conditioned on survival of the system. Simulation results are from $100$ realizations of the binary agreement dynamics. (b) Movement of the stable fixed point and the saddle point in phase space as a function of committed fraction $p$ (see text). The location of these points in phase space was obtained through a stability analysis (not shown) of the mean field equations Eq.~(\ref{MF}). The inset shows the density of nodes in state $B$ at the stable (red) and unstable(blue) fixed points as $p$ is varied; the expressions from which these numerical values are obtained are given in the Appendix. }
\label{Fig1}
\end{figure*}

The fixed-point and stability analyses (see Appendix) of these {\it mean-field} equations show that for any value of $p$, the consensus state in the committed opinion ($n_A = 1-p$, $n_B = 0$) is a stable fixed point of the mean-field dynamics.  However, below $p = p_c = \frac{5}{2}-\frac{3}{2} \left(\sqrt[3]{5+\sqrt{24}}-1\right)^2-\frac{3}{2}\left(\sqrt[3]{5-\sqrt{24}}-1\right)^2$, two additional fixed points appear: one of these is an unstable fixed point (saddle point), whereas the second is stable and represents an {\it active} steady state where $n_A, n_B$ and $n_{AB}$ are all non-zero (except in the trivial case where $p=0$) . Fig.~\ref{Fig1}(a) shows (asterisks) the steady state density of nodes in state $B$ obtained by numerically integrating the mean-field equations at different values of the committed fraction $p$ and with initial condition $n_A = 0, n_B = 1-p$. As $p$ is increased, the stable density of $B$ nodes $n_B$ abruptly jumps from $\approx 0.6504$ to zero at the critical committed fraction $p_c$. A similar abrupt jump also occurs for the stable density of $A$ nodes from a value very close to zero below $p_c$, to a value of $1$, indicating consensus in the $A$ state (not shown). In the study of phase transitions, an ``order parameter" is a suitable quantity changing (either continuously or discontinuously) from zero to a non zero-value at the critical point. Following this convention,  we use $n_B$ - the density of uncommitted nodes in state $B$ - as the order parameter appropriate for our case, characterizing the transition from an active steady state to the absorbing consensus state.

In practice, for a complete graph of any finite size, consensus is always reached. However, we can still probe how the system evolves, conditioned on the system not having reached consensus.  Fig.~\ref{Fig1}(a) shows the results of simulating the binary agreement model on a complete graph for different system sizes (solid lines). For $p < p_c$, in each realization of agreement dynamics, neglecting the initial transient, the density of nodes in state $B$, $n_B$, fluctuates around a non-zero steady state value, until a large fluctuation causes the system to escape from this active steady state to the consensus state. Fig.~\ref{Fig1}(a) shows these steady state values of $n_B$ conditioned on survival, for several values of $p$. As expected, agreement of simulation results with the mean-field curve improves with increasing system size, since Eq.~(\ref{MF}) represents the true evolution of the system in the asymptotic network size limit. Accordingly the critical value of the committed fraction obtained from the mean-field equations is designated as $p_c(\infty)$, although, for brevity, we refer to it simply as $p_c$ throughout this paper.

\begin{figure*}[!htbp]
\centerline{
\epsfxsize=16 cm
\epsfclipon
\epsfbox{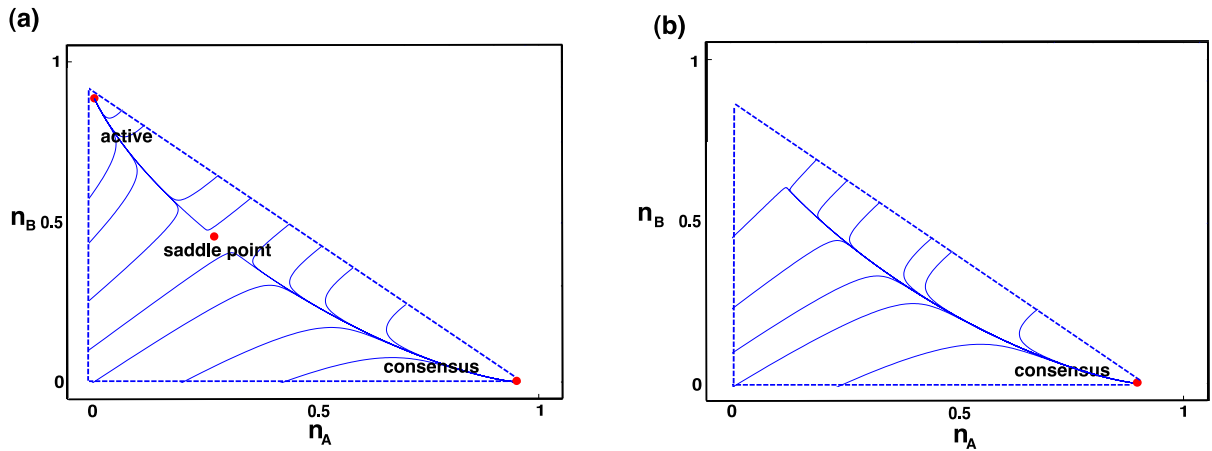}
}
\caption{Trajectories (obtained from integration of the mean-field equations Eq.~(\ref{MF})) in the phase-plane show the nature of flows from different regions of the phase-plane into existing fixed points for a) $p = 0.05$ ($<p_c$) and b) $p = 0.1$ ($>p_c$) }
\label{Fig2}
\end{figure*}

The existence of the transition as $p$ is varied and when the initial condition for densities is ($n_A=0,n_B=1-p $) can be further understood by observing the motion of the fixed points in phase space. Fig.~\ref{Fig1}(b) shows how the stable fixed point and the unstable fixed point move in phase space as $p$ is varied from $0$ to $p_c$. The active steady state moves downward and right while the saddle point moves upwards and left. At the critical value $p_c$ the two meet and the only remaining stable fixed point is the consensus fixed point. A similar observation was made in the model studied in \cite{Galam2007}. The fact that the value of $n_B$ converges to  $\approxeq 0.65$ and does not smoothly approach zero as the stable fixed point and the saddle point approach each other, explains the origin of the {\it first-order} nature of the phase transition. Fig ~\ref{Fig2} shows the representative trajectories obtained by integrating the mean-field equations for the cases where $p = 0.05$ ($< p_c$) and $p = 0.1$ ($> p_c$). 

\subsection{Finite network size: Scaling results for consensus times}
Even though consensus is always reached for finite $N$, limits on computation time prohibit the investigation of the consensus time, $T_c$, for values of $p$ below or very close to $p_c$.  We therefore adopt a semi-analytical approach prescribed in \cite{Dickman2002} that allows us to estimate the consensus times for different $N$ for an appreciable range of $p$ including values below $p_c$. We start with the master equation which describes the evolution of the probability that the network of size $N$ has $n$ ($m$) uncommited nodes in state $A$ ($B$). We denote by $c$, the number of committed nodes, and by $l (= N-n-m-c)$, the number of uncommitted nodes in state $AB$.
\begin{eqnarray}
\frac{dp_{nm}}{dt} \frac{1}{N} &=& \frac{1}{N^2} \bigg(-p_{nm} \; (2ln + \frac{3}{2} lc + 2nm + l(l-1) \notag \\ &+& 2lm + mc) + p_{n-1,m} \frac{3 (l+1)(n-1+c)}{2} \notag \\
   			     &+& p_{n+1,m}  \frac{(n+1)(2m + l-1)}{2} \notag \\  &+& p_{n-2,m} \frac{(l+2)(l+1)}{2} \notag \\ &+& p_{n,m-1}\frac{3(l+1)(m-1)}{2} \notag \\
			     &+& p_{n,m+1} \frac{(m+1)(2n+2c+l-1)}{2} \notag \\ &+& p_{n,m-2} \frac{(l+2)(l+1)}{2}\bigg)
\label{mastereq}			     
\end{eqnarray}
The factor of $1/N$ in the LHS comes from the fact that a transition between states takes place in an interval of time $1/N$. The transition rates in each term are the product of two densities, which is responsible for the overall factor of $1/N^2$ in the RHS. The probabilities are defined over all allowed states of the system ( i.e. $0 \leq n \leq N-c$, and $ 0 \leq m \leq N-c - n$ for given $n$  ) and the allowed transitions from any point $\{n,m\}$ in the interior of this state-space are $\{n,m\} \rightarrow \{n,m \pm 1\} , \{n,m\} \rightarrow \{n\pm1,m\}, \{n,m\} \rightarrow \{n,m+2\}, \{n,m\} \rightarrow \{n+2,m\}$.

We know from the mean field equations that in the asymptotic limit, and below a critical fraction of committed agents, there exists a stable fixed point. For finite stochastic systems, escape from this fixed point is always possible, and therefore it is termed {\it metastable}. For a finite system, the probability of having escaped to the metastable fixed point as a function of time is $P_e(t) = 1 - P_s(t)$ where $P_s(t)$ is the survival probability. The surviving fraction is constrained to be in the allowed region of $n,m$ quadrant excluding the true fixed point $\{N-c,0\}$. If the number of committed agents is far lower than $p_c~ N$ we expect that this surviving fraction will occupy configurations around the metastable fixed point, and the occupation probabilities $p_{n,m}$ will be peaked around the metastable fixed point. In systems which exhibit such long lived metastable states in addition to an absorbing fixed point, applying a quasistationary (QS) approximation has been found to be useful in computing quantities of interest \cite{Dickman2002_2,Dickman2002,Dallasta2006_2}. This approximation assumes that after a short transient, the occupation probability, conditioned on survival, of allowed states excluding the consensus state, is stationary. Following this approximation, the distribution of occupation probabilities conditioned on survival can be written as,  $\tilde{p}_{nm} = p_{nm}(t)/P_s(t)$ \cite{Dickman2002} and using this form in the master equation (Eq.~(\ref{mastereq})), we get:
\begin{eqnarray}
\frac{dP_s(t)}{dt} \tilde{p}_{nm} &=& - \frac{P_s(t)}{N} \bigg( \tilde{p}_{nm} \; (2ln + \frac{3}{2} lc + 2nm \notag \\ &+& l(l-1) +  2lm + mc) \notag \\ &-& \tilde{p}_{n-1,m} \frac{3 (l+1)(n-1+c)}{2} \notag \\
   			     &-& \tilde{p}_{n+1,m}  \frac{(n+1)(2m + l-1)}{2} \notag \\  &-&  \tilde{p}_{n-2,m} \frac{(l+2)(l+1)}{2} \notag \\ &-& \tilde{p}_{n,m-1}\frac{3(l+1)(m-1)}{2} \notag \\
			     &-& \tilde{p}_{n,m+1} \frac{(m+1)(2n+2c+l-1)}{2} \notag \\ &-& \tilde{p}_{n,m-2} \frac{(l+2)(l+1)}{2} \bigg)	
\label{mastereq_qs}     
\end{eqnarray}
Considering transitions from states $\{N-c-1,0\}$ and $\{N-c-2,0\}$ to the absorbing state $\{N-c,0\}$, we obtain the decay rate of the survival probability $dP_s(t)/dt$:
\begin{equation}
\frac{dP_s(t)}{dt} =  -P_s(t) \left[ \tilde{p}_{N-c-1,0} \left( \frac{3 (N-1)}{2N} \right) + \tilde{p}_{N-c-2,0} \frac{2}{N}  \right]
\label{survival}
\end{equation}
Substituting Eq.~(\ref{survival}) into Eq.~(\ref{mastereq_qs}), we finally obtain a condition that the occupation probabilities conditioned on survival must satisfy \cite{Oliveira2004} :
\begin{equation}
\tilde{p}_{nm} = \frac{\tilde{Q}_{nm}}{ W_{nm} - \tilde{Q}_0}
\label{QSeq}
\end{equation}
where, $\tilde{Q}_{nm} = Q_{nm}(t)/P_s(t)$ is obtained through explicit consideration of the terms in the master equation:
\begin{eqnarray}
\tilde{Q}_{nm} &=&  \tilde{p}_{n-1,m} \frac{3 (l+1)(n-1+c)}{2} \notag \\ &+&  \tilde{p}_{n+1,m}  \frac{(n+1)(2m + l-1)}{2}  + \tilde{p}_{n-2,m} \frac{(l+2)^2}{2} \notag \\
			   &+& \tilde{p}_{n,m-1}\frac{3(l+1)(m-1)}{2} + \tilde{p}_{n,m-2} \frac{(l+2)^2}{2} \notag \\ &+& \tilde{p}_{n,m+1} \frac{(m+1)(2n+2c+l-1)}{2}	
\label{Q}			   
\end{eqnarray}
and  $\tilde{Q}_0 = \left[ \tilde{p}_{N-c-1,0} \left( \frac{3 (N-1)}{2} \right) + 2~ \tilde{p}_{N-c-2,0}  \right]$ is the term arising from the decay of the survival probability (Eq.~(\ref{survival})). $W_{nm}$ is the coefficient of ${p}_{nm}$ ($\tilde{p}_{nm}$) within the brackets on the right hand side of Eq.~(\ref{mastereq}) (Eq.~(\ref{mastereq_qs})) and is equal to the transition rate out of state $\{n,m\}$ times $N^2$.

Eq.~(\ref{survival}) indicates that the survival probability decays exponentially with a rate  $\lambda = \tilde{Q}_0/N $. Since the mean lifetime of an exponentially decaying process is the inverse of the decay rate, it follows that the mean consensus time (neglecting the short transient before the QS distribution is attained) is
\begin{equation}
T_c \approxeq \frac{1}{\lambda} = 1/ \left[ \tilde{p}_{N-c-1,0} \left( \frac{3 (N-1 )}{2N} \right) + \tilde{p}_{N-c-2,0} \frac{2}{N}  \right]
\label{Tc}
\end{equation}

Thus, knowledge of $\tilde{p}_{nm}$s (in particular, $\tilde{p}_{N-c-1,0}$ and $\tilde{p}_{N-c-2,0} $) would allow us to calculate $T_c$ through Eq.~(\ref{Tc}). In order to obtain $\tilde{p}_{nm}$ (for all $0 \leq n,m$), we adopt the iterative procedure proposed in \cite{Oliveira2004}. Following this procedure, we start with an arbitrary initial distribution $\tilde{p}_{nm}^0$, and obtain a new distribution using: $\tilde{p}_{nm}^{i+1} = \alpha~\tilde{p}_{nm}^i + (1-\alpha)~\frac{\tilde{Q}_{nm}^{i}}{ W_{nm}^{i} - \tilde{Q}_0^{i}}$, where $0 \leq \alpha \leq 1$ is an arbitrary parameter, and $\tilde{Q}_{nm}^{i}, W_{nm}^{i}$ and $\tilde{Q}_0^{i}$ are all obtained using the probability distribution at the current iteration, $\tilde{p}_{nm}^i$ . With sufficient number of iterations, this procedure is expected to converge to a distribution that satisfies Eq.~(\ref{QSeq}) and which is thus, the desired QS distribution. In our case, we obtained acceptable convergence with a choice of $\alpha = 0.5$ and $30000$ iterations.

Following the above method, we obtain the QS distribution, and consequently the mean consensus times $T_c$ for different values of committed fraction $p$ and system size $N$. Fig.~\ref{Fig3}(a) shows how the consensus time grows as $p$ is decreased beyond the asymptotic critical point $p_c$ for finite $N$.  For $p<p_c$, the growth of $T_c$ is exponential in $N$ (Fig.~\ref{Fig3}(b), consistent with what is known regarding escape times from metastable states. For $p > p_c$, the QS approximation does not reliably provide information on mean consensus times, since consensus times themselves are small and comparable to transient times required to establish a QS state. However, simulation results show that above $p_c$ the scaling of mean consensus time with $N$ is logarithmic (Fig.~\ref{Fig3}(c)). A snapshot of the QS distribution (Fig.~\ref{Fig4}) near $p_c$ ($p = 0.09$) for a system of size $N=100$ shows clearly the bimodal nature of the distribution, with the two modes centered around the stable fixed point, and the consensus fixed point.

The precise dependence of consensus times on $p$ can also be obtained for $p < p_c$ by considering the rate of exponential growth of $T_c$ with $N$. In other words, assuming $T_c \sim \exp(\alpha(p)N)$, we can obtain $\alpha(p)$ as a function of $p$. Figure ~\ref{Fig3}(d) shows that $\alpha(p) \sim |p-p_c|^{\nu}$ where $\nu \approx 1.65$. Thus, below $p_c$, we have:
\begin{equation}
T_c(p < p_c) \sim \exp((p_c-p)^\nu N)
\end{equation}
This exponential growth is presumably modulated by factors of $\log N$ which become dominant only when $p = p_c$. Above $p_c$, the dependance of $T_c$ on $p$ as seen from simulations is negligible (not shown).

\section{Sparse Networks}
Next, we present simulation results for the case when the underlying network topology is chosen from an ensemble of Erd\H{o}s-R\'enyi (ER) random graphs with given size $N$ and given average degree $\kav$. The qualitative features of the evolution of the system in this case are the same as that of the complete graph, although the critical fraction $p_c$ displays some dependence on $\kav$. For small $\kav$ and fixed $N$, the drop in consensus times occurs slightly earlier in $p$ for ER graphs than for a complete graph of the same size, as shown in Fig.~\ref{Fig5}(a). However for $p > p_c$, a complete graph has shorter consensus times (on average) than an ER graph of the same size. Above $p_c$, the difference between consensus times for a graph with average degree $\kav$ and the complete graph, $\Delta T_c$, decays approximately as power law with increasing $\kav$ (Fig.~\ref{Fig5}(b)). The deviation from a perfect power law is likely due to other weaker $\kav$ dependent terms, presumably logarithmic in $\kav$. 
\begin{figure*}[!htbp]
\centerline{
\epsfxsize=16cm
\epsfbox{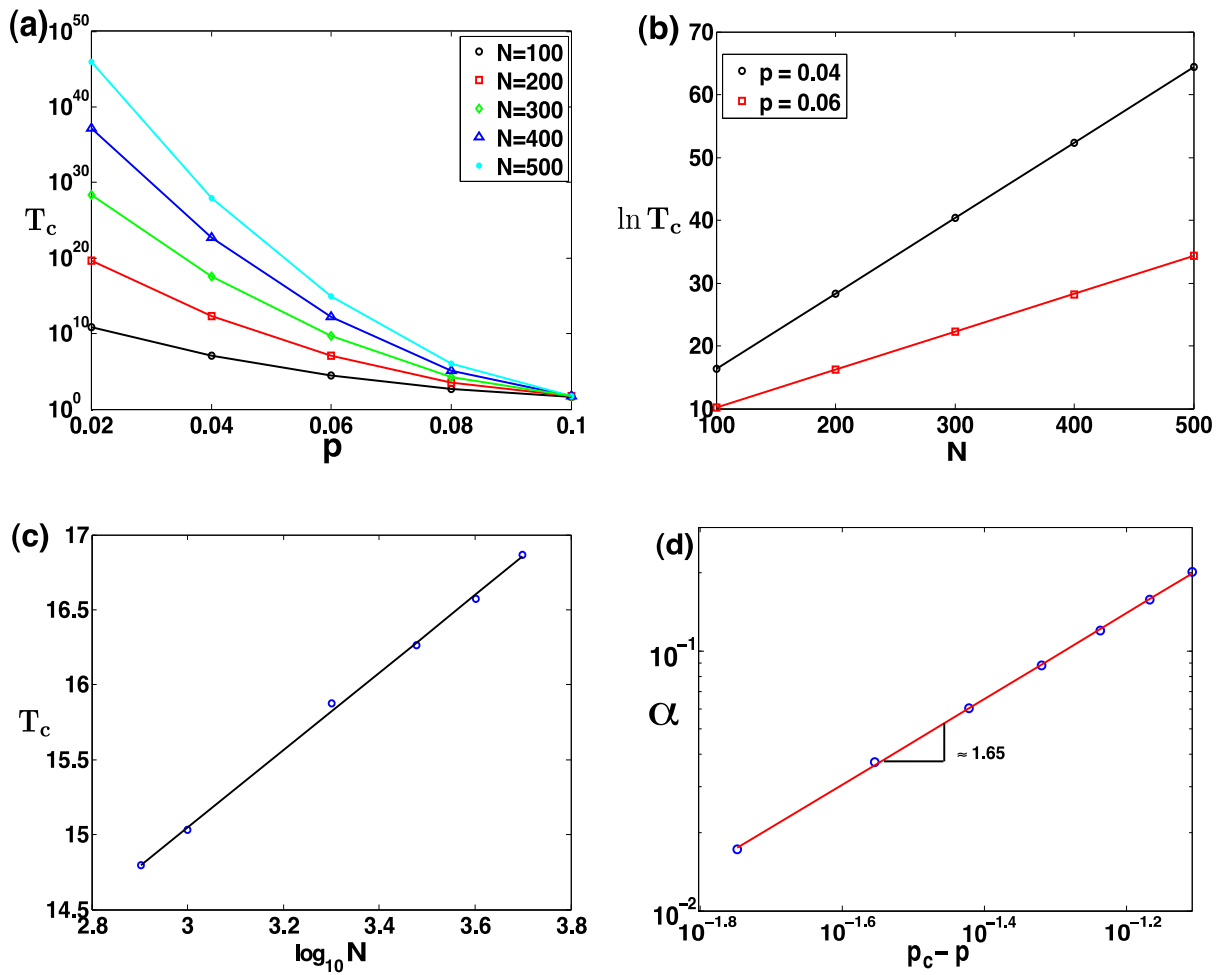}
}
\caption{(a) Mean consensus time $T_c$ for $p < p_c$ obtained by using the QS approximation. (b) Exponential scaling of $T_c$ with $N$, for $p < p_c$; mean consensus times (circles, squares) are obtained using the QS approximation. The lines are guides to the eye. (c) Logarithmic scaling of $T_c$ with $N$ for $p = 0.3 > p_c$; mean consensus times are obtained from simulations. The line shows the best linear fit to the data. (d) The rate, $\alpha(p)$, of exponential growth of the consensus time with $N$ as a function of $p-p_c$ (see text). Circles show the values of $\alpha(p)$ obtained for $p = 0.2,0.3,0.4,0.5,0.6,0.7,0.8$ by considering the scaling of $T_c$ with $N$, for these values of $p$. The straight line shows a linear fit to the data plotted on a log-log scale. }
\label{Fig3}
\end{figure*}
We also performed simulations of the binary agreement model on Barabasi-Albert (BA) networks (Fig.~\ref{Fig3}c)), and found similar qualitative behavior  as observed for ER networks including the difference from mean-field behavior. We leave a detailed analysis of the dependence of the critical fraction $p_c$ and the consensus times $T_c$ on the average degree $\kav$ of sparse networks, for future work.
\begin{figure}[!htbp]
\centerline{
\xsize
\epsfclipon
\epsfbox{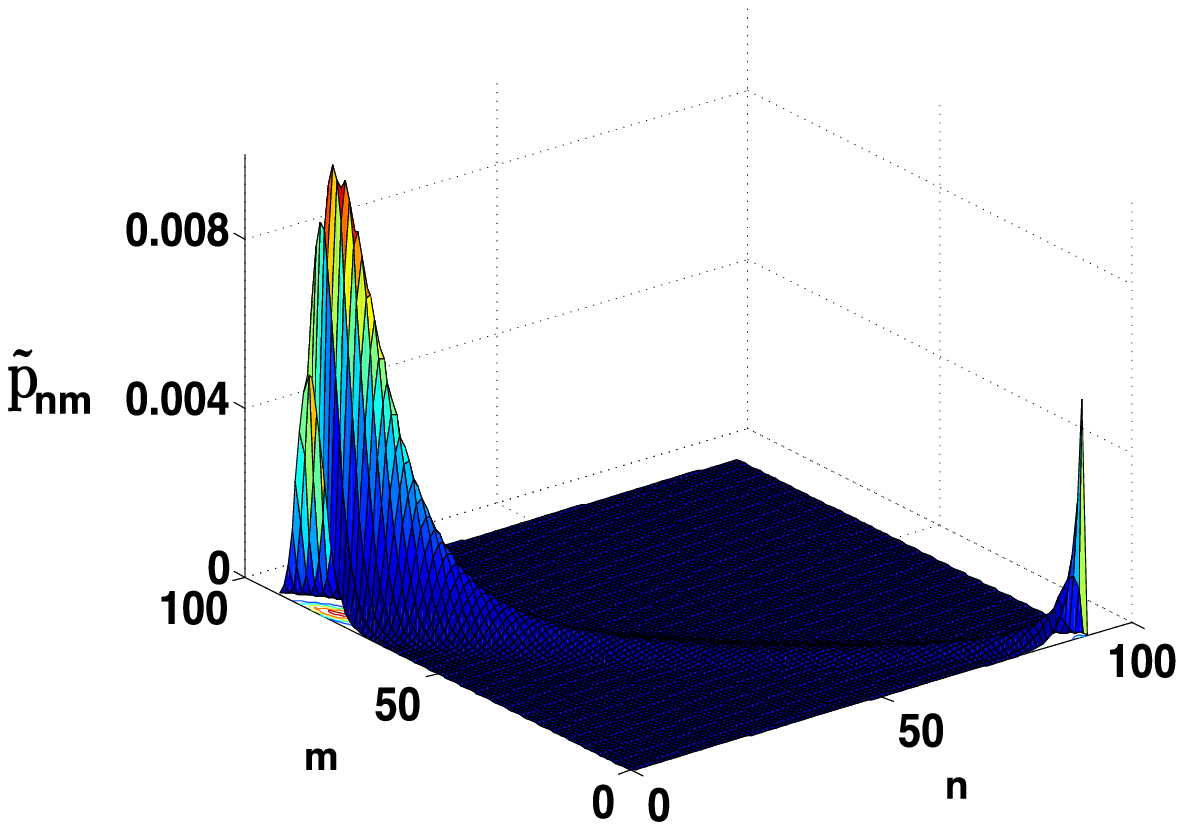}
}
\caption{ The quasistationary distribution $\tilde{p}_{nm}$ for $p = 0.09$ and $N = 100$.  }
\label{Fig4}
\end{figure}

\section{Summary}
In closing, we have demonstrated here the existence of a {\it tipping point} at which the initial majority opinion of a network switches quickly to that of a consistent and inflexible minority. There are several historical precedents for such events, for example, the suffragette movement in the early 20th century and the rise of the American civil-rights movement that started shortly after the size of the African-American population crossed the $10 \%$ mark. Such processes have received some attention in sociological literature under the term {\it minority influence} \cite{Moscovici1969, Galam2007}. Our motivation here has been to study this process in more detail through semi-analytical methods and simulations for finite-sized and sparse networks, within the realm of a particular social influence model - the binary agreement model. There are several open questions and extensions of this work that are worth studying, in our opinion: for example, given a network with non-trivial community structure, what is the optimal scheme for selecting committed agents (for a given committed fraction) that would minimize consensus times, and reduce $p_c$? Secondly, extensions of the model to include utility-driven opinion switching by agents may be useful in designing optimal incentive schemes for opinion spreading.
\begin{figure}[htbp]
\centerline{
\epsfxsize=8cm
\epsfclipon
\epsfbox{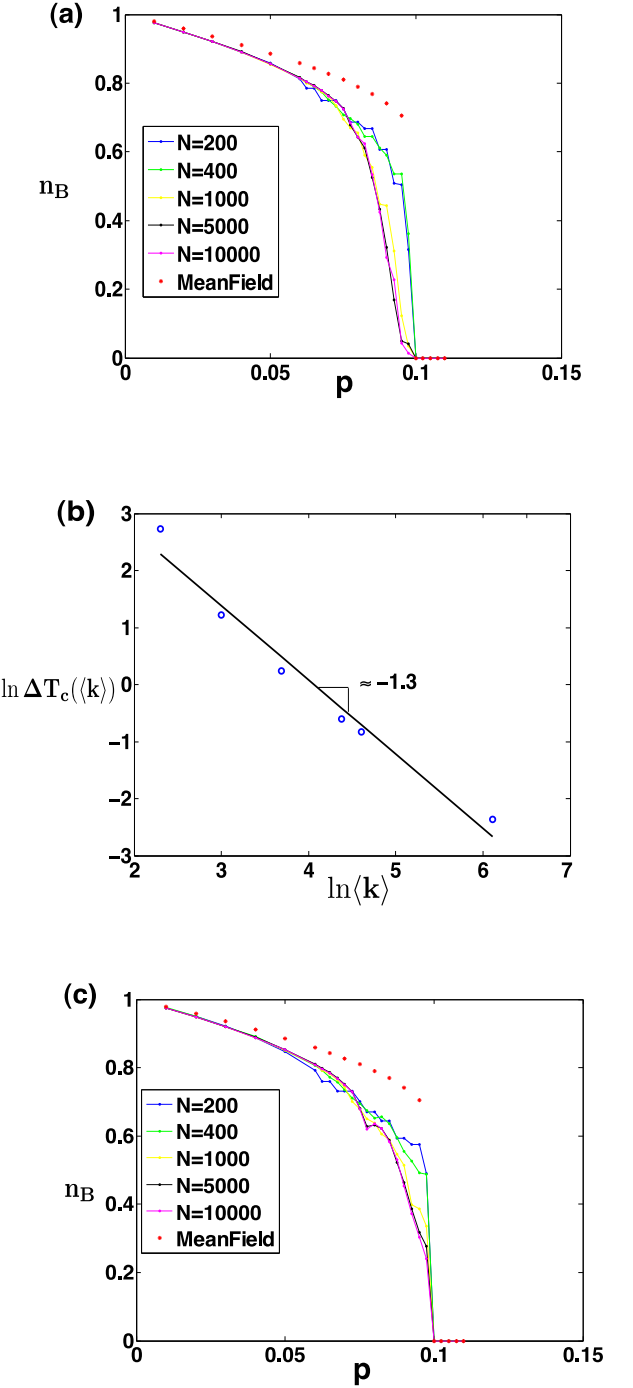}
}
\caption{(a) The steady state density $n_B$ of nodes in state $B$ as a function of committed fraction $p$ for Erd\H{o}s-R\'enyi graphs  of different sizes with $\kav = 10$, conditioned on survival of the system. Lines are mean values of $n_B$ obtained from simulations; asterisks show mean-field consensus times obtained by integrating Eq. ~\ref{MF} (b) Scaling of $\Delta T_c(\kav)$ (defined in text) with $\kav$; the line shows the best linear fit to the data. (c) The steady state density $n_B$ of nodes in state $B$ as a function of committed fraction $p$ for Barabasi-Albert (BA) networks of different sizes with $\kav = 10$, conditioned on survival of the system}
\label{Fig5}
\end{figure}

\section{Acknowledgments}
This work was supported in part by the Army Research Laboratory under Cooperative Agreement Number W911NF-09-2-0053, by the Army Research Office grant W911NF-09-1-0254 and by the Office of Naval Research Grant No. N00014-09-1-0607. The views and conclusions contained in this document are those of the authors and should not be interpreted as representing the official policies either expressed or implied of the Army Research Laboratory or the U.S. Government. S.S. thanks R. Dickman and M. M. de Oliveira for useful discussions.
  
\section*{Appendix: Fixed points of the mean-field equations}
 Here, we analyze the mean-field equations for the existence of fixed points. To simplify notation we use the notation $x = n_A$ and $y = n_B$. Thus for a fixed point of the evolution given by Eq.~(\ref{MF}):
\begin{eqnarray}
-xy &+& (1-x-y-p)^2+x(1-x-y-p) \notag \\ &+& 1.5p(1-x-y-p) = 0 \notag \\
 		-xy &+& (1-x-y-p)^2+y(1-x-y-p) \notag \\&-&yp  =  0 
\label{FP}
\end{eqnarray}
which can be reduced further to:
\begin{eqnarray}
		x & = & [(1-y-p/4)^2-9p^2/16]/(p/2+1)  \notag \\
 		y& = & (1-x-p)^2
 \label{eq:fixpt2}
\end{eqnarray}
Substituting the expression of $x$ into the expression for $y$, and denoting $z^2=y$ we get
\begin{equation}
	z(z^3- (2-p/2)z+p/2+1)=0
 \label{eq:fixpt}
\end{equation}
For any value of $p$, $z  = z_0 = 0 $ is a solution to the above equation. In other words, for any value of $p$, the mean-field equations admit a stable fixed point, $n_A = x_1 = 1-p$, $n_B = y_1 = 0$ which represents the network having reached a consensus state where all nodes have adopted the opinion of the committed agents. 

The remaining fixed points are roots of 
\begin{equation}
f(z) = z^3- (2-p/2)z+p/2+1 = 0.
\label{fz}
\end{equation}
In order to find the criterion which has to be satisfied for valid roots (i.e. $0 \leq z \leq \sqrt((1-p))$) of the above equation to exist, we analyze the extrema of the function $f(z)$ which are given by:
\begin{equation}
	f'(z)=3z^2-2+p/2=0 
\label{derivative}
\end{equation}
Thus, the extrema occur at :
\begin{equation}
	z_{1,2}=\pm \sqrt{2/3-p/6}
\label{extrema}
\end{equation}
It can be seen from Eq.~(\ref{derivative}) that $f(z)$ is increasing, decreasing and increasing again in the intervals $(- \infty,z_1)$,$(z_1,z_2)$,$(z_2,+ \infty)$ respectively. Consequently, $f(z)$ achieves a maximum at $-1<z_1=-\sqrt{2/3-p/6}<0$ and a minimum at $0<z_2=\sqrt{2/3-p/6}<1$. Furthermore, since $f(-2)=-p/2-3<0$ and $f(-1)=2>0$, one root of $f(z) = 0$ occurs in the interval $-2 < z < -1$. Since $f(z)$ is positive at $z_1$, decreasing from $z_1$ to $z_2$ where it achieves a minimum, and increasing thereafter, it follows that a necessary and sufficient condition for more roots of $f(z) = 0$ to exist, is that $f(z_2) $ be less than zero:
\begin{equation*}
	f(z_2)=z_2^3-(2-p/2)z_2+p/2+1<0 
\end{equation*}
Denoting $z_2=q$ and $p=4-6q^2$ (from Eq.~(\ref{extrema})) yields the following inequality for $q$ as a condition for more roots of $f(z) = 0$ to exist:
\begin{equation}
	f(q)=q^3+1.5q^2-1.5>0 \label{eq:fixpt5}
\end{equation}
(Note that $z_2$ is itself a function of $p$).
Analyzing the derivative of $f(q)$ enables us to glean that the inequality Eq.~(\ref{eq:fixpt5}) is satisfied for $q > q_0$ where $q_0$ is the solution of the cubic equation $f(q) = 0$ and is given by:
\begin{equation*}
	q_0=\left [ \sqrt[3]{5+\sqrt{24}}+ \sqrt[3]{5-\sqrt{24}} -1 \right ]/2
\end{equation*}
Thus, the original fixed point equation Eq.~(\ref{FP}) has at least one valid root besides $z=0$, so long as $p$ is less than or equal to:
\begin{equation}
	p_c=\frac{5}{2}-\frac{3}{2} \left(\sqrt[3]{5+\sqrt{24}}-1\right)^2-\frac{3}{2}\left(\sqrt[3]{5-\sqrt{24}}-1\right)^2
\end{equation}
which using standard computer algebra software is evaluated to be $p_c = 0.09789$. Using, $z^2 = y = q_0$ and Eq.~(\ref{eq:fixpt2}), we obtain the state of the system at $p_c$ to be $\{n_A,n_B\} = \{ 0.0957, 0.6504\}$.
It also follows from the expression for $f(z)$, that $f(0) > 0$ and therefore if $f(z_2)$ is negative, Eq.~(\ref{fz}) has {\it two} roots on the positive line when $p < p_c$. Thus there are two fixed points of Eq.~(\ref{FP}) when $p < p_c$.

The exact expressions for these fixed points (that can also be obtained numerically), obscure their dependence on $p$. We therefore adopt an approximation which exhibits a much clearer dependence of the fixed point values on $p$, and numerically yield values close to those obtained from the exact expressions. Substituting $z = t\sqrt{ 2-\frac{p}{2}}$ in Eq.~(\ref{fz}) reduces it to:
\begin{equation}
t^3 - t + r = 0
\label{cubic}
\end{equation}
where $r = \frac{1+p/2}{(\sqrt{2-p/2})^3}$. Clearly, $r$ is a monotonically increasing function of $p$, and therefore $\frac{1}{2\sqrt{2}} \leq r < \frac{1+p_c/2}{(\sqrt{2-p_c/2})^3}=\frac{2}{3\sqrt{3}}$ for $0 \leq p < p_c$, our range of interest. Function $g(t)=t^3-t$ is monotonically decreasing for $t<-1$ and $g(-1)=0$, while $g(-2)<-1<-r$. Hence, there is a real root $t_1\in(-2,-1)$ to Eq.~(\ref{cubic}) which is a monotonically decreasing function of $r$, but which clearly does not yield a valid fixed point. This root can be expressed as $t_1(r)=-\frac{2}{\sqrt{3}}+\epsilon(r)$, where $\epsilon(r)$ is monotonically decreasing from less than $0.0106$ to $0$ over the range of our interest for $r$. Substituting this expression back into Eq. \ref{cubic} and neglecting powers of $\epsilon(r)$ higher than unity, we get an approximation of $\epsilon$ in terms of $r$, and consequently an approximation for $t_1$:
\begin{equation}
t_1(r) \approx -\frac{16}{9\sqrt{3}} - \frac{r}{3}  
\end{equation}
with relative error of less than 0.01\%.
Now, we can factorize the LHS of Eq.~(\ref{cubic}) and write it as $(t^2 + bt + c) (t - t_1)$. Equating this factorized expression with $t^3 - t + r$, gives us $b$ and $c$ in terms of $r$.
Thus, two more roots of Eq.~(\ref{cubic}) are obtained in terms of $r$ by solving the quadratic equation $t^2 + bt + c = 0$ which yields:
\begin{equation}
t_{2,3} =  \frac{8}{9\sqrt{3}} + \frac{r}{6} \pm \sqrt{\frac{17}{81} - \frac{8r}{9\sqrt{3}} -\frac{r^2}{12}}
\label{t_roots}
\end{equation}
Finally, we can obtain the values of $z$ associated with the above roots, and therefore the values of $x$ and $y$ written in terms of these roots are derived as:
\begin{eqnarray}
y_{2,3} &=& t_{2,3}^2 \frac{4-p}{2} \notag \\
x_{2,3} &=& \frac{(4- 4y_{2,3} -p)^2 - 9p^2}{8p + 16}
\label{msvalues}
\end{eqnarray}


The stability of these fixed points can be checked via linear stability analysis. Following the standard procedure, the stability matrix is given by:
\[
S = 
\left[ \begin{array}{cc}
-1-\frac{p}{2} & -2+2y^*+\frac{p}{2} \\
-2+2x^*+2p & -1\\
\end{array}
\right]
\]
where $(x^*,y^*)$ is the fixed point under consideration. The eigenvalues of the stability matrix are given by:
\begin{eqnarray}
\lambda = \frac{1}{4} \bigg( -4 - p \pm ( 17p^2 &+& 64 (x^*-1)(y^*-1) \notag \\ &+&16p(x^*+4y^*-5) )^\frac{1}{2} \bigg)
\label{eigenvalues}
\end{eqnarray}
From the expression for the eigenvalues we numerically determine that the real part of both the eigenvalues is negative for $(x_2,y_2)$ over the range $0 \leq p < p_c$ indicating that $(x_2,y_2)$  that  is a stable fixed point. This is however, not the case for $(x_3,y_3)$, making it unstable. Similarly, the consensus fixed point $(x_1,y_1)$ is found to be stable for $0 \leq p \leq 1$.
Finally, we note that as $p \to 0$, the stable fixed point converges to $n_A = 0,n_B=1$, while the unstable fixed point converges to $n_A = n_B \approxeq 0.38$. 


\def\urlprefix{}
\def\url#1{}



\end{document}